\documentclass[conference]{IEEEtran}

\usepackage{cite}      

\usepackage{graphicx}  

\usepackage{url}       
\begin{document}

\title{Open-architecture Implementation of\\
Fragment Molecular Orbital Method\\
for Peta-scale Computing}

\author{\authorblockN{Toshiya Takami\authorrefmark{1},
Jun Maki\authorrefmark{1},
Jun-ichi Ooba\authorrefmark{1},
Yuichi Inadomi\authorrefmark{2},
Hiroaki Honda\authorrefmark{2},
Taizo Kobayashi\authorrefmark{1},\\
Rie Nogita\authorrefmark{1},
and Mutsumi Aoyagi\authorrefmark{1}\authorrefmark{2}}
\authorblockA{\authorrefmark{1}Computing and Communications Center,
Kyushu University, 6--10--1 Hakozaki, Higashi-ku, Fukuoka 812-8581, Japan}
\authorblockA{\authorrefmark{2}PSI Project Laboratory, Kyushu University,
3--8--33--710 Momochihama, Sawara-ku, Fukuoka 814-0001, Japan}
}


%


\maketitle

\begin{abstract}
We present our perspective and goals
on high-performance computing for nanoscience
in accordance with the global trend toward ``peta-scale computing.''
After reviewing our results obtained through
the grid-enabled version of the fragment molecular orbital method (FMO)
on the grid testbed by the Japanese Grid Project,
National Research Grid Initiative (NAREGI),
we show that FMO is one of the best candidates for peta-scale applications
by predicting its effective performance in peta-scale computers.
Finally, we introduce our new project constructing
a peta-scale application in an open-architecture implementation
of FMO in order to realize both goals of
high-performance in peta-scale computers
and extendibility to multi-physics simulations.
\end{abstract}


%
\IEEEpeerreviewmaketitle

\section{Introduction}

On account of the recent developments of the grid computing environment,
we can use many computer resources more than a thousand CPUs.
However, those large distributed resources are accessible
only when we pass through some gateway to the grid system
after a tedious procedure for grid-enabling of application programs.
Thus, for scientists in nanoscience to use the grid system
as one of the daily tools,
it is important to make their applications grid-aware in advance.

On the other hand, the development on
high performance computing (HPC) environments shows no sign of slowing down,
and, within several years, we might reach the scale of peta ($\sim10^{15}$)
in computer resources for scientific computing.
It is expected that the ``peta-scale computer''
exhibits more than a peta flops performance in floating-point calculations,
its available memory is more than a peta byte in total,
or it has external storages amount to more than a peta byte.
Thus, the global trend in HPC is now to study
how to realize the peta-scale computing\cite{JSPS}.

The purpose of this paper is twofold:
one is to present our computational results in nanoscience
supported by the Japanese Grid project,
National Research Grid Initiative (NAREGI)\cite{NAREGI-Web},
and the other is to show a perspective about
the HPC applications for peta-scale computing.
This paper is coordinated as follows.
Those computational results
using the fragment molecular orbital method (FMO)
on grid computing environments
are shown in section \ref{sec:NAREGI}.
The performance prediction of the FMO calculation
in a peta-scale computer is shown in section \ref{sec:prediction-peta},
and the introduction of the actual implementation of FMO
in order to realize peta-scale computing
is given in section \ref{sec:open-fmo}.
Finally in section \ref{sec:discussion},
we discuss what is necessary in the actual peta-scale computing
for scientific applications.

\section{Grid-enabled FMO}
\label{sec:NAREGI}

In this section,
we review the results on a grid-enabled version of FMO,
and evaluate its effectivity in the grid testbed of NAREGI\cite{NAREGI}.
Our main contribution to the project is to develop
large-scale grid-enabled applications in nanoscience.
One of these applications is the Grid-FMO
which is based on the famous {\it ab initio\/} molecular
orbital (MO) package program, GAMESS\cite{GAMESS},
and is constructed by dividing the package
into several independent modules
so that they are executed on a grid environment.
The algorithm of FMO and
the implementation of Grid-FMO is briefly reviewed
in the following subsections.

\subsection{Algorithm of FMO}
\label{sec:fmo-algorithm}

\begin{figure}
\begin{center}
\includegraphics[scale=0.45]{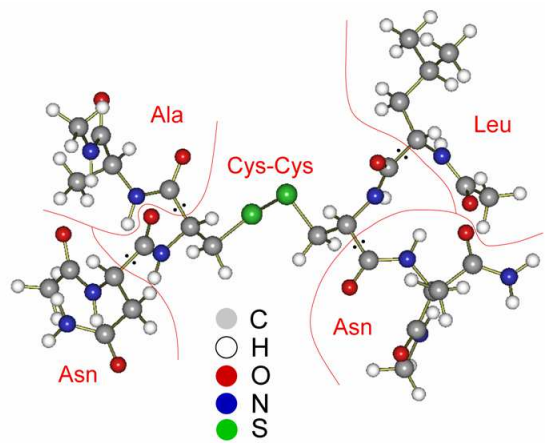}
\caption{In FMO, the target molecule is divided into fragments
for which {\it ab initio\/} calculation is performed.
Usually, carbon--carbon single bonds are chosen as a boundary
between fragments.}
\end{center}
\end{figure}

The FMO method developed by Dr. Kitaura and co-workers\cite{FMO}
can execute all electron calculation in
large molecules with more than 10 thousands atoms.
The brief overview of the flow of the fragment MO method
up to dimer correction is the following:
(1) the molecule to be calculated is divided into fragments,
(2) {\it ab initio\/} MO calculation\cite{Roothaan} is executed
for each fragment (monomer) under the electro-static potential
made by all the other fragments, and this calculation is
repeated until the potential is self-consistently converged,
(3) {\it ab initio\/} MO calculation is executed
for pairs of fragments (dimer),
and (4) the total energy is obtained by summing up all the results.
This algorithm of FMO is implemented in several famous MO packages
such as GAMESS\cite{GAMESS}, ABINIT-MP\cite{ABINIT-MP}, etc.

Although the most time-consuming part of this calculation is
{\it ab initio\/} MO calculations for each fragment
and each pair of fragments,
these calculations can be executed independently.
Thus, FMO is easily executed in parallel computers with high efficiency.
The FMO routines included in those famous packages are already parallelized
for cluster machines, and are coordinated
so that they exhibit efficient performance.
However, if the program is used in distributed computing environments,
we should consider robustness and controllability
in addition to the efficiency.
Thus, grid-enabling is necessary.
Among many ways of grid-enabling, we choose a strategy
to reconfigure the program into a loosely-coupled form
in order to satisfy such properties.

\subsection{Loosely-coupled FMO}

\begin{figure}
\begin{center}
\includegraphics[scale=0.43]{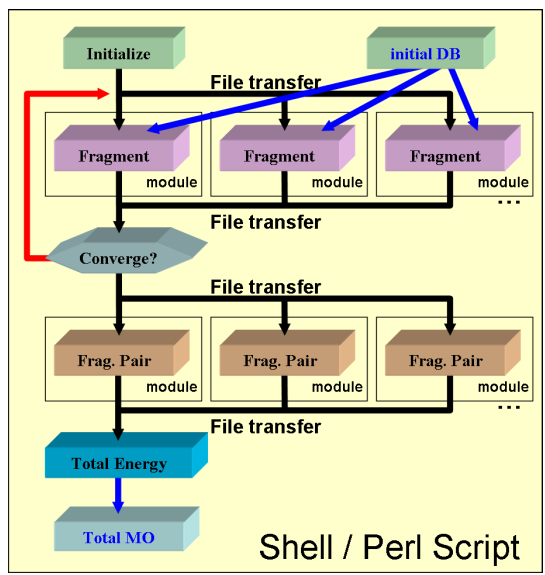}
\caption{Flow of calculation of the grid-enabled FMO}
\label{fig:lc-fmo-flow}
\end{center}
\end{figure}

\begin{figure}
\begin{center}
\includegraphics[scale=0.43]{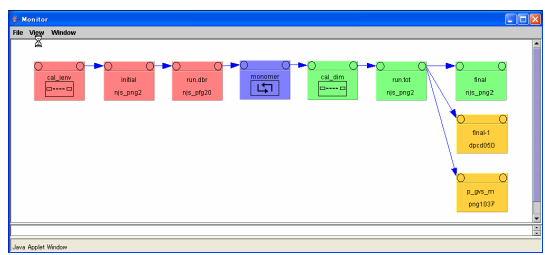}
\caption{Flow of LC-FMO represented in NAREGI Workflow Tool}
\label{fig:wft}
\end{center}
\end{figure}

We have developed a grid-enabled version of FMO,
called a Loosely-coupled (LC) FMO program\cite{LC-FMO},
as part of the NAREGI project.
At first, we show the procedure to change the GAMESS-FMO programs
into a ``loosely-coupled form,''
where the original FMO in the GAMESS package
is divided into several ``single task'' modules
which are connected each other through input/output file transfers.
Those modules consists of
{\tt run.ini} for the initialization,
{\tt run.mon} for the fragment calculation (monomer),
{\tt run.dim} for the fragment pair calculation (dimer),
and {\tt run.tot} for the total energy calculation,
and are invoked from a script program which also manages
the file transfer over distributed computers.
The total flow of the LC-FMO is shown in fig. \ref{fig:lc-fmo-flow}.

Another benefit of the loosely-coupled form is
the extendibility of its functionality.
In fact, after the first release of the LC-FMO,
we have developed two other modules
which realize a linkage to the initial density database
and a visualization feature of the total molecular orbitals.
Since the top-level program to invoke modules
is lightweight and can be written in script languages,
further modification of the total flow is easily performed.

\subsection{Execution on NAREGI Grid}

In order to execute LC-FMO on the NAREGI grid,
we put the flow of the program into NAREGI Workflow Tool\cite{NAREGI}.
The graphical representation of the LC-FMO program
is shown in fig. \ref{fig:wft},
where modules and input/output files are represented
in icons connected each other.
This procedure is simple and straightforward
because we have already reconstructed the program
into a suitable form for distributed computing.
Thus, the important is whether the basic structure of the program
is grid-aware or not.

Once a program is grid-enabled,
we can execute it by the use of the large-scale computing resources
over one thousand CPUs.
From the programmer's points of view,
the most preferable thing is that we are relieved of
arranging the resources to execute the program in high efficiency.
In NAREGI testbed system\cite{NAREGI},
NAREGI Super Scheduler can manage the resource rearrangement
with the help of NAREGI Information Service.

\subsection{Efficiency of LC-FMO}

Robustness and efficiency are conflicting each other in general.
Since our reconstruction of FMO to increase the controllability
might hurt the efficient execution of the program,
it is necessary to evaluate the efficiency of LC-FMO
by comparing with the original FMO program in GAMESS.

\begin{figure}
\begin{center}
\includegraphics[scale=0.45]{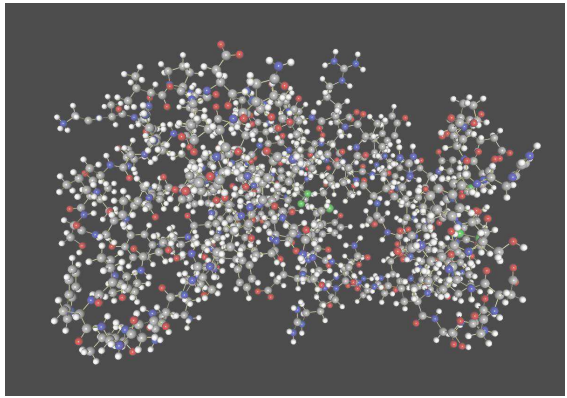}
\caption{The molecular structure of chicken egg white cystatin
(1CEW, 1701 atoms, 106 fragments).}
\label{fig:lc-fmo-benchmark}
\end{center}
\end{figure}

In table \ref{tbl:lc-fmo-elapsed},
we show elapsed times for chicken egg white cystatin (1CEW, 1701 atoms,
106 fragments, shown in fig.~\ref{fig:lc-fmo-benchmark})
by the LC-FMO and the original GAMESS-FMO codes.
This is obtained on the NAREGI testbed system
which consists of 16 CPUs of Xeon 3GHz.
Generally speaking, an increase of the total elapsed time is inevitable
when we reconstruct some application into a loosely-coupled form,
which is considered as a cost for grid-enabling.
As shown in table \ref{tbl:lc-fmo-elapsed},
however, the increase of the time is relatively small.
Thus, it is evaluated that the LC-FMO is effectively
executed on grids.

\begin{table}
\caption{The elapsed time for chicken egg white cystatin.}
\label{tbl:lc-fmo-elapsed}
\begin{center}
\begin{tabular}{l|rr}\hline
	& LC-FMO & GAMESS-FMO \\ \hline\hline
Initial Guess & 37s & 4s \\ \hline
Monomer & 1h11m & 59m \\ \hline
Dimer & 2h16m & 2h11m \\ \hline
Energy & 4s & $<$ 1s \\ \hline\hline
Total & 3h28m & 3h10m \\ \hline
\end{tabular}
\end{center}
\end{table}

\section{Performance Prediction}
\label{sec:prediction-peta}

In this section, we predict what performance would be expected
if we executed our FMO program in a ``peta-scale computer.''
Since the actual hardware is not available at present,
the prediction is based on
a hypothetical specifications of the computer.

\subsection{Hypothetical Specification of the Peta-scale Computer}

It is said that, in a few years, such computer systems with
peta-scale specifications will be designed\cite{petaComputer}.
Here, we concentrate on the CPU resources of the peta-scale computer,
and estimate how many CPU resources are necessary
to attain peta-scale performances.

The current peak performance of a single-core CPU is
the order of 10 giga flops.  If a parallel computer
with 10 peta-flops peak performance is configured by the CPU
to achieve 1 peta-flop effective performance,
1,000,000 CPU cores are necessary in total.
Since cluster system constructed by 1,000,000 machines
connected each other is unimaginable,
it is necessary to rearrange those resources into multiple layers.
We do not concern here how to realize computational nodes
with multiple CPU cores.
If we can configure a lowest layer by a node
with the performance of 100 $\sim$ 1,000 CPUs,
the total computer will be a parallel system
composed of 1,000 $\sim$ 10,000 nodes.

\subsection{Performance Prediction of FMO}

The procedure to predict the performance of FMO
in the peta-scale computer is a somewhat phenomenological method
based on actual measurements in current computer systems.
First, we assume an execution model of FMO
which gives a theoretical timing function
of the number of fragments $N_f$.
After fixing some parameters by several measurements
of the program, we predict the elapsed time
on the virtual computer with peta-scale specifications.

\subsubsection{Execution Model of GAMESS-FMO}

\begin{table}
\caption{Timing data obtained in a single node of IBM p5 1.9GHz.\label{tbl:ibm}}
\begin{tabular}{l|l||r|r|r|r}
\multicolumn{2}{l||}{Input}
	& 1cew	& 1ao6\_half & 1ao6 & 1ao6\_dim \\ \hline\hline
\multicolumn{2}{l||}{No. Atom}
	& 1701	& 9121	& 18242	& 36484	\\
\multicolumn{2}{l||}{$N_f$}
	& 106	& 561	& 1122	& 2244	\\
\multicolumn{2}{l||}{$N_d(N_f)$}
	& 690	& 4192	& 8416	& 16832	\\
\multicolumn{2}{l||}{$N_{es}(N_f)$}
	& 4875	& 152888 & 620465 & 2499814 \\
\multicolumn{2}{l||}{$I_m$}
	& 17	& 17	& 17	& 17	\\ \hline
	& monomer	& 1356	& 13364	& 40005	&140810	\\
	& (Average)	&(0.752)&(1.40)	&(2.10)	&(3.69)	\\ \cline{2-6}
Time	& SCF-dimer	& 2037	& 20689	& 70465	&186901	\\
(sec)	& (Average)	&(2.95)	&(4.94)	&(8.37)	&(11.1)	\\ \cline{2-6}
	& ES-dimer	& 398	& 13772	& 55955	&208627	\\
	& (Average)	&(0.0816)&(0.0901)&(0.0902)&(0.0835)\\
\hline
\multicolumn{2}{l||}{Elapsed Time (sec)}
	& 3799&47886&166601&536898
\end{tabular}
\end{table}

\begin{table}
\caption{Timing data obtained in 16 CPUs of Xeon 3GHz.\label{tbl:xeon}}
\begin{tabular}{l|l||r|r|r}
\multicolumn{2}{l||}{Input}
	& 1cew	& 1ao6\_half & 1ao6 \\ \hline\hline
\multicolumn{2}{l||}{No. Atoms}
	& 1701	& 9121	& 18242	\\
\multicolumn{2}{l||}{$N_f$}
	& 106	& 561	& 1122	\\
\multicolumn{2}{l||}{$N_d(N_f)$}
	& 690	& 4192	& 8416	\\
\multicolumn{2}{l||}{$N_{es}(N_f)$}
	& 4875	& 152888 & 620465 \\
\multicolumn{2}{l||}{$I_m$}
	& 17	& 17	& 17	\\ \hline
	& monomer	& 1030.4 &10808.9 &33989.2	\\
	& (Average)	& (9.72) & (19.3) & (30.3)	\\ \cline{2-5}
Time	& SCF-dimer	& 1677.4 &17517.6 &50819.2	\\
(sec)	& (Average)	& (41.3) & (71.0) & (102.7)	\\ \cline{2-5}
	& ES-dimer	& 293.1  & 9594.8 &39133.4	\\
	& (Average)	& (1.02) & (1.07) & (1.07)	\\ \hline
\multicolumn{2}{l||}{Elapsed Time (sec)}
	& 3003.5&38065.9&126330.9
\end{tabular}
\end{table}

Even if you could analyze all the program codes of GAMESS,
it is almost impossible to determine the number of floating-point
calculations in a FMO routine precisely
since there are many conditional branches
which depend on the molecular structure given as an input.
Our strategy, here, is a practical approach
to obtain phenomenological values of the execution time
through experimental executions of FMO in the current machines.

The total execution time of FMO is divided into three parts:
a monomer part, an SCF-dimer part, and an ES-dimer part.
The monomer and SCF-dimer parts perform SCF calculations
for each fragments and each pair of fragments, respectively,
while the ES-dimer part obtains dimer correction terms
under an electro-static (ES) approximation between fragments.
Usually, the ES approximation is applied to the fragment pair
with fragments which are separated more than a certain threshold.
We start from an expression of the total amount of computation
as a function of the number of fragments $N_f$,
\begin{equation}
  F_{\rm total}(N_f)=F_m(N_f)+F_d(N_f)+F_{es}(N_f),
\end{equation}
where $F_m(N_f)$, $F_d(N_f)$, and $F_{es}(N_f)$
represent the number of floating-point calculations
for monomer, SCF-dimer, and ES-dimer parts, respectively.
These are assumed as
\begin{eqnarray}
\label{eqn:Fm}
  F_m(N_f)&=&\left[f_m^{(0)}+f_m^{(1)}N_f\right]N_fI_m\\
\label{eqn:Fd}
  F_d(N_f)&=&\left[f_d^{(0)}+f_d^{(1)}N_f\right]N_d(N_f)\\
\label{eqn:Fes}
  F_{es}(N_f)&=&f_{es}^{(0)}N_{es}(N_f),
\end{eqnarray}
according to the algorithm of the FMO theory\cite{FMO},
where $I_m$ is the number of monomer loops,
$N_d(N_f)$ and $N_{es}(N_f)$ represent
the numbers of SCF-dimers and ES-dimers.
SCF calculations for monomers and dimers in FMO
depend on the environmental potential which is made by
all the fragments other than the target monomer and dimer,
while it can be ignored for ES-dimers.
Thus, we represent the amount of each SCF calculation
for a monomer and an SCF-dimer up to a linear term of $N_f$,
and that of non-SCF calculation for an ES-dimer as a constant.
The parameters, $f_m^{(0)}$, $f_m^{(1)}$, $f_d^{(0)}$, $f_d^{(1)}$,
$f_{es}^{(0)}$, should be determined from the actual timing data.
If we define additional parameters,
the number of parallel nodes $K$ and the effective floating-point
performance $E$ (flops) of each node,
we can compare the actual timing data with the amount of computation
divided by $KE$.

\begin{figure}
\begin{center}
\includegraphics[scale=0.5]{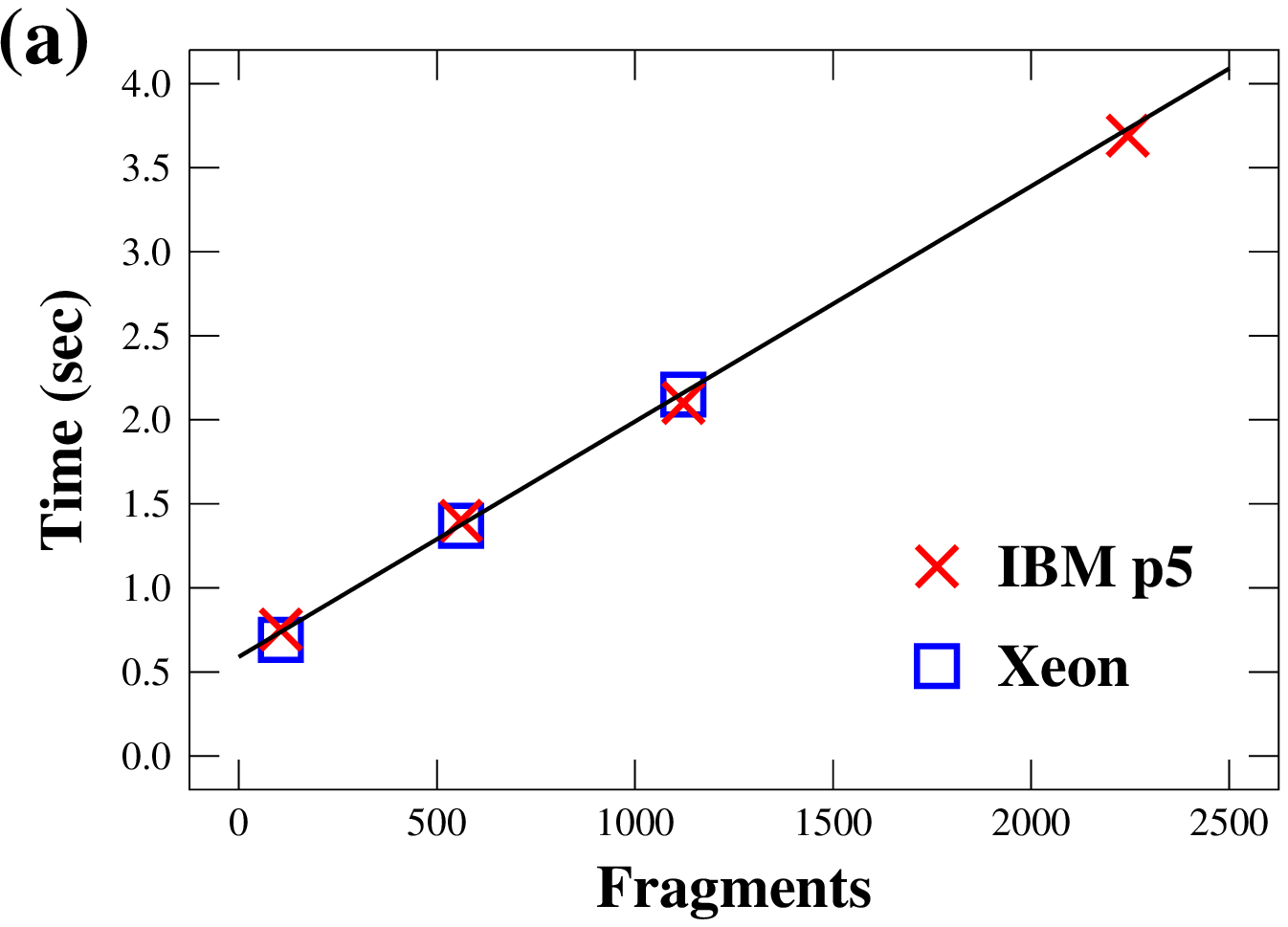}\\
\includegraphics[scale=0.5]{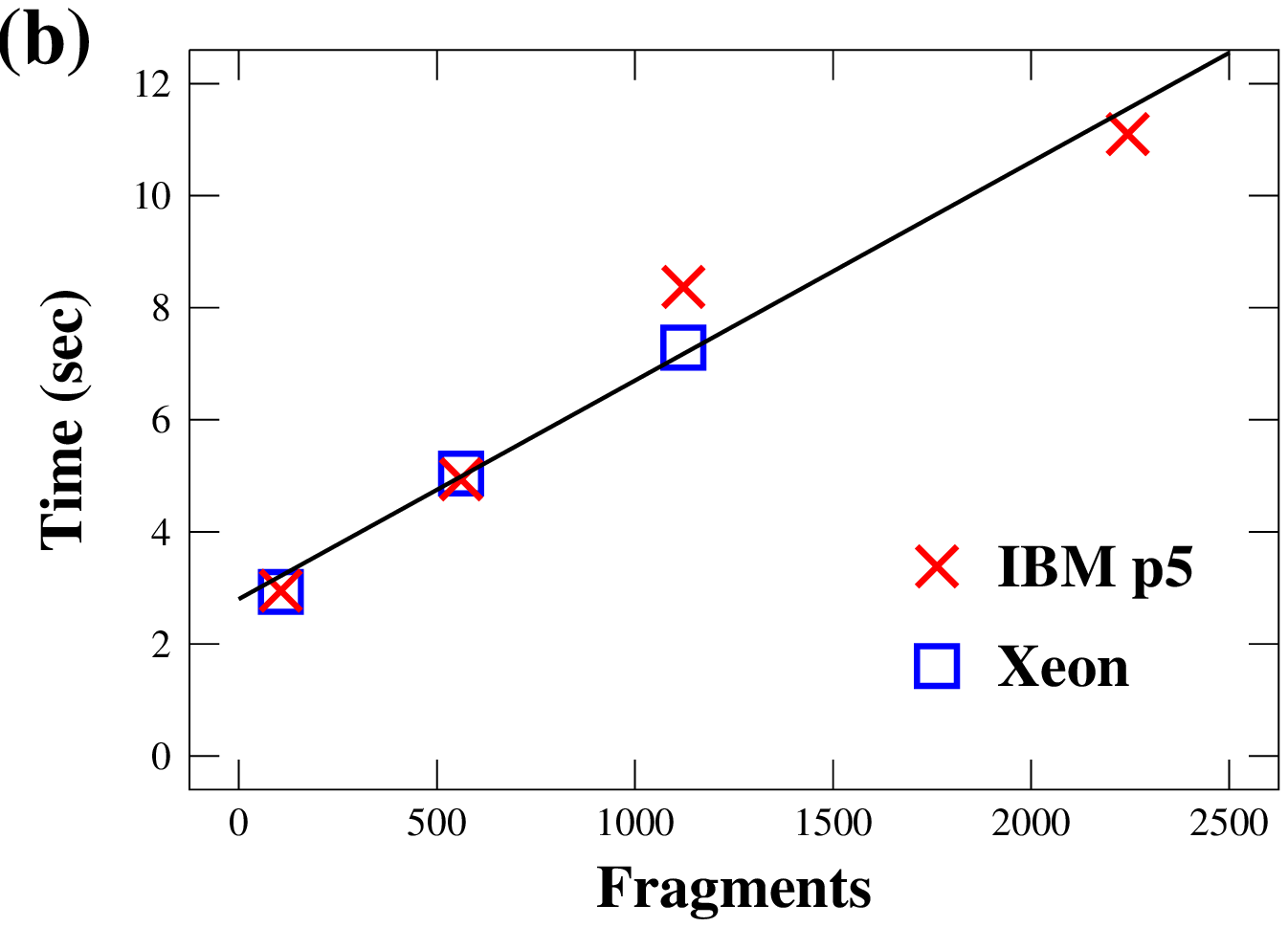}\\
\includegraphics[scale=0.5]{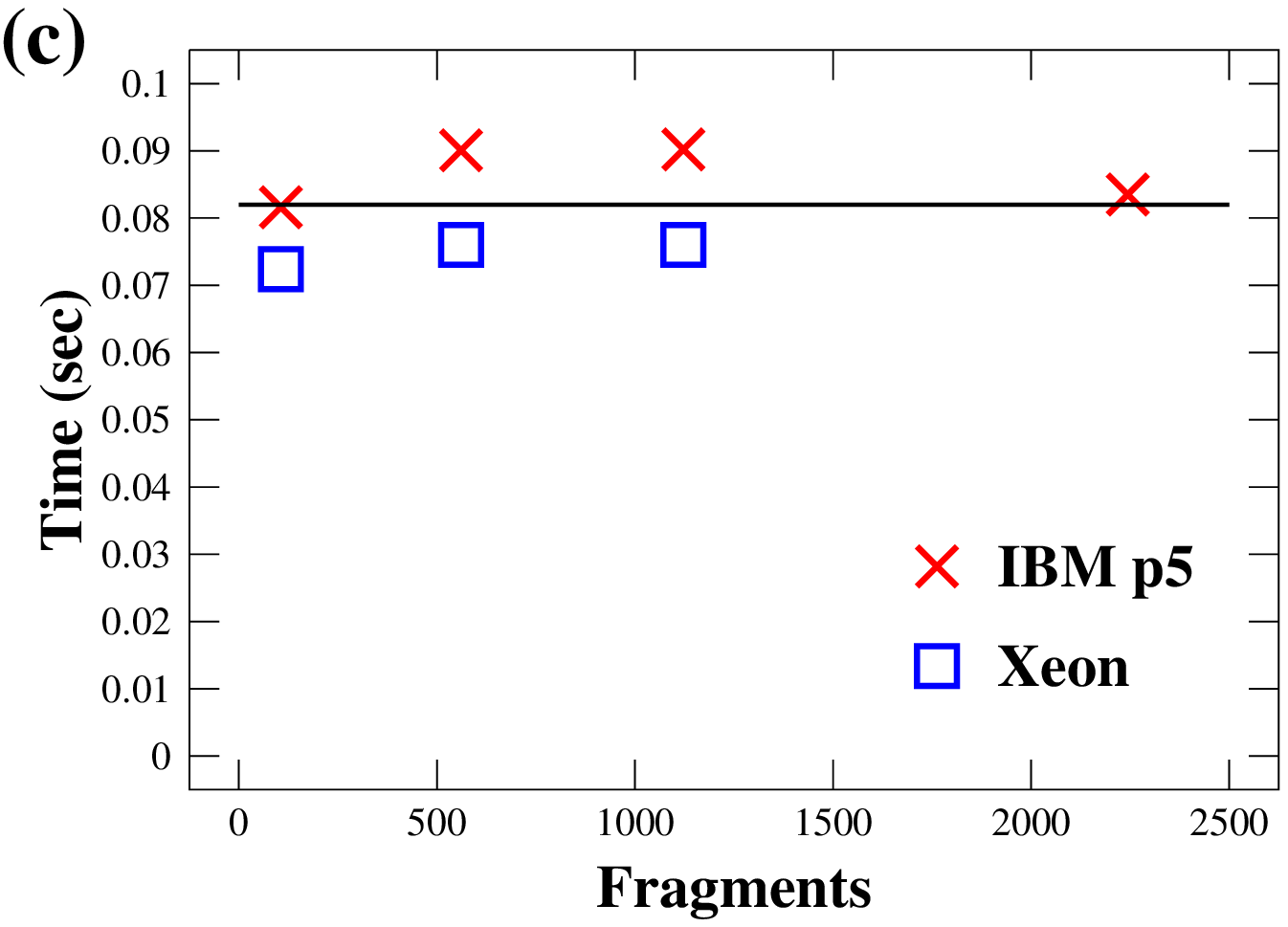}
\caption{The fragment number dependency
of the amount of computation for
(a) a monomer, (b) an SCF-dimer, and (c) an ES-dimer
are plotted with the effective performance ratios
($E=1$ for IBM p5, $E=0.071$ for Xeon).
Lines are functions (a) $(f_m^{(0)}+f_m^{(1)}N_f)/KE$,
(b) $(f_d^{(0)}+f_d^{(1)}N_f)/KE$,
and (c) $f_{es}^{(0)}/KE$ obtained by the least-square method.
\label{fig:time}
}
\end{center}
\end{figure}

Table \ref{tbl:ibm} and \ref{tbl:xeon} show the timing data of each part
in test executions of several inputs on the machines,
1 node of IBM p5 1.9GHz and a 16 CPU cluster of intel Xeon 3GHz.
Since the averaged size of fragments in these inputs are considered
as almost the same, the difference of the elapsed time
mainly depends on the number of the total fragments $N_f$.
The least-square fitting with the additional parameters
representing the effective performances of each computer,
$E_{\rm ibm}$ and $E_{\rm xeon}$,
determines all the parameter values shown in table \ref{tbl:parameter}
after the normalization $E_{\rm ibm}=1$.
In figures \ref{fig:time} (a), (b) and (c),
we plot the timing data and the results of the functions,
(a) $(f_m^{(0)}+f_m^{(1)}N_f)/KE$, (b) $(f_d^{(0)}+f_d^{(1)}N_f)/KE$,
and (c) $f_{es}^{(0)}/KE$.

\begin{table}
\begin{center}
\caption{Parameter values to represent the execution model of FMO
\label{tbl:parameter}}
\begin{tabular}{c|c||l|l}
\multicolumn{2}{c||}{Parameter} & \multicolumn{2}{c}{Value} \\ \hline\hline
$f_m^{(0)}$ & $f_m^{(1)}$ & 0.59 & 0.0014 \\ \hline
$f_d^{(0)}$ & $f_d^{(1)}$ & 2.83 & 0.0039 \\ \hline
$f_{es}^{(0)}$ & --- & 0.082 & --- \\ \hline\hline
$E_{\rm ibm}$ & $E_{\rm xeon}$ & 1.0 & 0.071
\end{tabular}
\end{center}
\end{table}

\begin{figure}
\begin{center}
\includegraphics[scale=0.5]{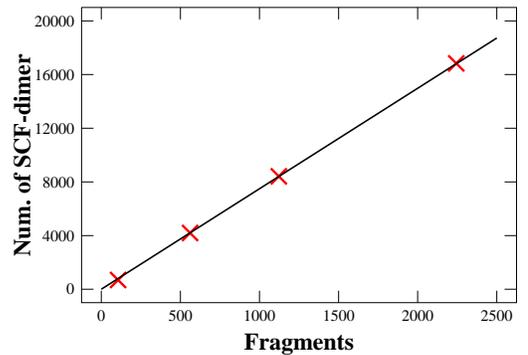}
\caption{The number of SCF-dimers in the test executions
and the function $N_d(N_f)$ by the least-square method.}
\end{center}
\end{figure}

\subsubsection{Performance in Peta-scale Computer}

In order to predict the total performance of FMO,
it is convenient if the functions $N_d(N_f)$ and $N_{es}(N_f)$ are
represented in simple functions of $N_f$.
By the least-square fitting of the data
in table \ref{tbl:ibm} and \ref{tbl:xeon},
we obtain a function for the number of SCF-dimers
\begin{equation}
  N_d(N_f)=7.50\ N_f.
\end{equation}
If we consider a constraint
\begin{equation}
  N_d(N_f)+N_{es}(N_f)=\frac{(N_f-1)N_f}2,
\end{equation}
the number of ES-dimers is represented in the form
\begin{equation}
  N_{es}(N_f)=\frac{(N_f-1)N_f}2-7.50\ N_f.
\end{equation}

\begin{figure}
\begin{center}
\includegraphics[scale=0.5]{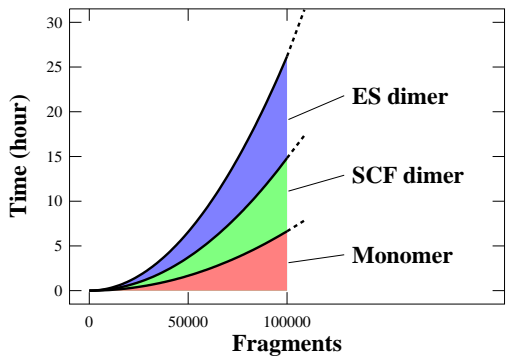}
\caption{\small
\label{fig:peta}
Predicted elapsed time of the FMO calculation by GAMESS
as a function of the number of fragments $N_f$.
This is obtained on the assumption that
10,000 nodes are available, and each node is 5 times
faster than a current machine.}
\end{center}
\end{figure}

If we substitute these functions and parameters
into Eqs. (\ref{eqn:Fm}), (\ref{eqn:Fd}), and (\ref{eqn:Fes}),
the total computational amount of FMO is obtained
as a function of $N_f$.
As already seen in the previous section,
the FMO algorithm is appropriate for parallel executions
when the number of fragments is large enough
compared to the number of available nodes.
Suppose that we have a peta-scale computer
with $K=10000$ and $E=5$, i.e.,
the number of available nodes is 10,000 and each node
has an effective speed 5 times faster than a node of IBM p5 1.9GHz
which is used in this paper as a reference machine ($E_{\rm ibm}=1$).
Then, the predicted elapsed time $F_{\rm total}(N_f)/KE$ is
calculated as a quadratic function of $N_f$ shown in fig. \ref{fig:peta}.
From the viewpoint of the elapsed time,
we can perform quantum calculations for
molecules with more than 100,000 fragments
if the peta-scale computer is realized.

According to the effective performance measurements on PCs,
The FMO calculation is executed in 0.5 $\sim$ 1.0 giga flops
per one CPU of Xeon or Pentium4, which means that
our reference machine, a node of IBM p5 1.9GHz,
exhibits almost 10 giga flops for the program
since it is $E_{\rm ibm}/E_{\rm xeon}\approx14$ times faster than a Xeon.
Then, the total performance of FMO calculations
for a peta computer with $K=10000$ and $E=5$
is considered as 0.5 peta flops.
Thus, FMO is considered as a predominant candidate
which can record peta-scale performance.

\section{OpenFMO Project}
\label{sec:open-fmo}

In spite of the fact that the FMO is a suitable algorithm
to achieve the peta-scale performance through large-scale parallel executions,
the actual GAMESS-FMO code cannot be executed
for the molecule with more than 100,000 fragments.
One of the reason why the current program fails to run
is memory consumption in each node,
where the current FMO code in GAMESS
tries to allocate two-dimensional arrays with respect to fragments,
e.g., the distance between two fragments.
This exceeds 40 giga bytes even if we consider the symmetry.
The limit of the number of fragment is considered
as the order of 10,000 in the current implementation.
Thus, the reconstruction of the FMO code is necessary
in order to correspond to the peta-scale computing.

\begin{figure}
\begin{center}
\includegraphics[scale=0.48]{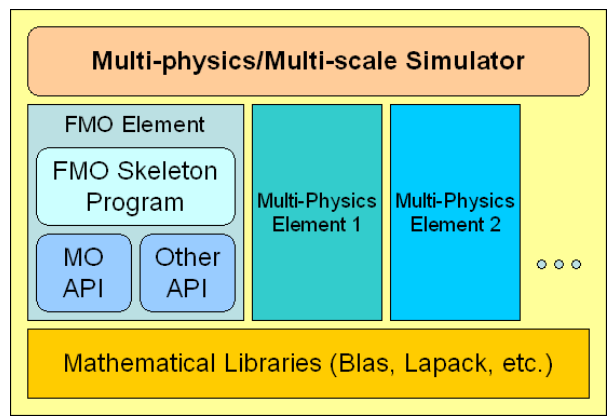}
\caption{Multi-physics/multi-scale simulator stack including OpenFMO.}
\label{fig:stack}
\end{center}
\end{figure}

\begin{figure}
\begin{center}
\includegraphics[scale=0.5]{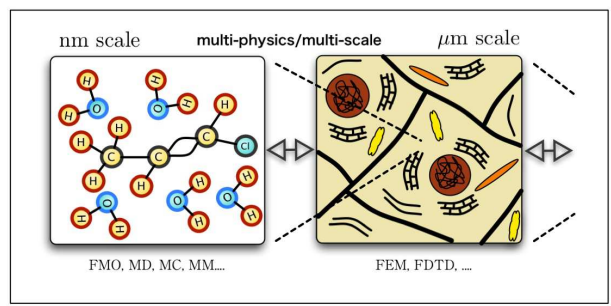}
\caption{Can multi-physics/multi-scale simulations
reveal the true aspect of the complex world of matter?}
\label{fig:multiscale}
\end{center}
\end{figure}

In this section, we introduce our new project
to reconstruct a FMO program from scratch.
The name of this project, OpenFMO\cite{OpenFMO},
stands for the following openness:
(1) names and argument lists of APIs constructing
the FMO program are opened publicly,
(2) the program structure is also opened to combining
other theories of physics and chemistry,
i.e., multi-scale/multi-physics simulations,
and (3) the skeleton program is developed under some open-source licenses
and its development process will also be opened to the public.
In fig. \ref{fig:stack}, we show a stack structure of this program.
Although the main target is still
a quantum chemical calculation of molecules,
it can be combined with other theories
to construct multi-physics/multi-scale simulations
(fig. \ref{fig:multiscale}) for complex phenomena\cite{OE-Analysis}.

\subsection{Open Architecture Implementation of FMO}

As already shown in section \ref{sec:fmo-algorithm},
the fragment MO method consists of the standard
{\it ab initio\/} MO calculations
including the generation of Fock matrices
with one/two-electron integral calculations.
Using this property, the usual FMO program is divided into two layers,
the skeleton program which control the whole flow of the FMO algorithm,
and the molecular orbital (MO) APIs to provide the charge distribution
of each fragment through {\it ab initio\/} MO calculations.
Since these spec of interfaces between the skeleton
and the APIs are fixed and opened publicly,
either of them can easily be substituted by other programs.

\subsection{Open Interface to Multi-physics Simulations}

The multi-physics simulation is one of the predominant strategy
to construct peta-performance applications for nano-scale materials.
Our new implementation of FMO can also be opened
for such multi-physics simulations.
Since the FMO method is based on
electro-static interaction between fragments,
we can extend each fragment to the general object
which can provide a static charge distribution.
For examples, we often use molecular mechanics representation
of atomic clusters which should be given by quantum mechanical description,
or the environmental charge distribution surrounding a molecule
which models a solvent can be included as a fragment.
Thus, the MO-APIs for a certain fragment can be substituted
to other programs based on the different approximation levels.

In addition to the description in molecular sciences,
this is extended to larger-scale simulations
through the ``multi-physics/multi-scale simulator'' layer.
This type of extension is widely used for the large-scale computation
representing realistic models of molecules in cells or other living organisms.

\subsection{Open Source Development of the Programs}

The source code of the skeleton program of OpenFMO is publicly opened
according to some open-source licenses.
At present, the OpenFMO project is managed by several core members
including the authors of this paper.
Once we show the effectiveness of our approach to the open implementation,
and the direction of the project is settled properly,
we are willing to change the management of the project
into so-called open-source-software developments.

\section{Summary and Discussion}
\label{sec:discussion}

In this paper, we showed our results in nanoscience
executed on the NAREGI grid system,
and expressed our perspective toward the peta-scale computing.
In section \ref{sec:prediction-peta},
we showed that FMO is one of the peta-scale applications.
However, it has been also realized
that the actual implementations of the FMO method, at present,
does not correspond to the execution in peta-scale environments,
i.e., they does not solve molecules with more than 10 thousand
fragments even if peta-scale computers are available.
In order to improve the situation,
we started an open-source project for multi-physics simulations
including new FMO codes from scratch.
\vfill\eject

Generally speaking, it is very difficult
for scientific applications to be executed in a peta-scale performance
since a simple parallel scheme fails in case that
the total number of CPUs is the order of 1,000,000.
Thus, it will be necessary that the application itself
becomes multi-layered in order to use the resources efficiently
corresponding to the hardware architecture of peta-scale computers.
The FMO calculation which we have studied in this paper
has a two-stage structure in its original algorithm:
{\it ab initio\/} calculations are performed
for each fragment and fragment pair while the interactions between
these fragments are described by a classical electro-static potentials.
This theoretical reconstruction of the original {\it ab initio\/} MO method
into the layered form is considered as the key
to the successful performance of the program.

Scientific calculations with layered structures have been carried out
as a multi-physics/multi-scale simulation in various fields.
This type of calculations is important
not only as a technique of realistic simulations
but also as an actual example for the peta-scale computing.
However, we claim that the deep understanding of the physical/chemical
theories must be necessary for a proper construction of effective
simulations which describes complex aspects of nature.
Then, the theoretical and practical way of constructing such simulations
should be established before the time when the peta-scale computers
are available.
We expect that our implementation of the simulator
is one of the solutions for high-performance scientific simulations
in the next-generation.


\section*{Acknowledgment}
This work is partly supported by the Ministry of Education, Sports, Culture,
Science and Technology (MEXT) through the Science-grid NAREGI Program
under the Development and Application of Advanced High-performance
Supercomputer Project.



%

\end{document}